\documentclass[conference]{IEEEtran}
\IEEEoverridecommandlockouts

\usepackage{cite}
\usepackage{amsmath,amssymb,amsfonts}
\usepackage{algorithmic}
\usepackage{graphicx}
\usepackage{textcomp}
\usepackage{xcolor}
\usepackage{stfloats}
\graphicspath{{../figs/}}
\usepackage{caption}
\usepackage{listings}
\usepackage{subcaption}
\DeclareGraphicsExtensions{.png}
\usepackage{float}
\def\BibTeX{{\rm B\kern-.05em{\sc i\kern-.025em b}\kern-.08em
    T\kern-.1667em\lower.7ex\hbox{E}\kern-.125emX}}

\begin{document}

\newcommand{\leo}[1]{\textcolor{red}{#1}}
\newcommand{\leon}[1]{\textcolor{red}{#1}}
\newcommand{\mikel}[1]{\textcolor{blue}{#1}}
\newcommand{\red }[1]{\textcolor{red}{#1}}

\author{\IEEEauthorblockN{Mikel Cortes-Goicoechea}
\IEEEauthorblockA{\textit{Barcelona Supercomputing Center} \\
Barcelona, Spain \\
mikel.cortes@bsc.es}
\and
\IEEEauthorblockN{Leonardo Bautista-Gomez}
\IEEEauthorblockA{\textit{Barcelona Supercomputing Center} \\
Barcelona, Spain \\
leonardo.bautista@bsc.es}
}

\title{Discovering the Ethereum2 P2P Network}


\maketitle
\thispagestyle{plain}
\pagestyle{plain}

\begin{abstract}

Achieving the equilibrium between scalability, sustainability, and security while keeping decentralization has prevailed as the target solution for decentralized blockchain applications over the last years. Several approaches have been proposed by multiple blockchain teams to achieve it, Ethereum being among them. Ethereum is on the path of a major protocol improvement called Ethereum 2.0 (Eth2), implementing Sharding and introducing the Proof-of-Stake (PoS). As the change of consensus mechanism is a delicate matter, this improvement will be achieved through different phases, the first of which is the implementation of the Beacon Chain. As Ethereum1, Eth2 relies on a decentralized peer-to-peer (p2p) network for the message distribution. Up to date, we estimate that there are around 5.000 nodes in the Eth2 main net geographically distributed. However, the topology of this one still prevails unknown. In this paper, we present the results obtained from the analysis we performed on the Eth2 p2p network. Describing the topology of the network, as possible hazards that this one implies. 

\end{abstract}


\section{Introduction}
\label{sec:intro}

Ethereum \cite{eth-whitepaper} is a mature decentralized web infrastructure that has proved reliability and had led the growth of decentralized applications. It features an ecosystem for decentralized applications based on a general purpose virtual machine~\cite{hildenbrandt2018kevm} and its dedicated programming language~\cite{wohrer2018smart}. This ecosystem holds certain characteristics, which has generated a large community of developers that have proposed new methods to tackle some of the limitations affecting the Ethereum protocol.

The Ethereum Foundation together with the Ethereum community have been working on the development of the second generation of Ethereum, Ethereum 2.0 (Eth2). The approach proposed in the Eth2 protocol tackles both: the scalability challenge by introducing sharding, and the sustainability challenge by changing the consensus protocol to Proof-of-Stake (PoS). This ambitious project is divided into different phases because of its complexity. The first phase (phase 0) is the Beacon Chain \cite{phase-0} launched on December 2020.

Eth2 aims to create a major network of nodes that will provide the infrastructure for sharding. The different developer teams have been working over the last two years to generate Eth2 client implementations, each client targeting a specific type of user. This variety of clients, together with the low computing-power required by the Eth2 protocol, offers to most of the users the possibility to join the network. 

In contrast to the Ethereum1 client, the Eth2 one is divided in two main parts, the Beacon Node and the Validator. The beacon node, or node, performs all kind of interactions with the Eth2 network. It establishes the connections with other nodes, exchanges the messages, downloads the blocks and maintains the view of the Beacon Chain as the view of the Shards that the validator is performing at, building among them the entire network. On the other hand, the validator is in charge of the logical part, generating the attestations on the received blocks, as well as proposing the blocks when needed. To perform its duties, the validator needs to follow the current state of the Chain, and to be in constant communication with a Beacon Node that provides such information. 

A project with this complexity level completely relies on the healthiness of the network. To propagate the messages between the nodes of the network, Eth2 relies on the GossipSub p2p protocol \cite{vyzovitis2019gossipsub}. The protocol is based on the exchange of messages and metadata between peers distributed in meshes, achieving the message propagation within the targeted time-margins and minimizing wasteful bandwidth consumption. Given the Eth2 network requirements and the interaction between the clients, it is important to monitor critical points for possible attacks. In this paper, we present a complete analysis of the p2p network of the Eth2 main net. The analysis offers a general overview of the network composition, becoming the first analysis on the recently launched Eth2 main net.

The remainder of this paper is organized as follows. 
Section~\ref{sec:background} discusses the background and related work.
Section~\ref{sec:stateofart} exposes the protocols involved on the Eth2 project.
Section~\ref{sec:methodology} explains the methodology used for our study and evaluation.
Section~\ref{sec:analysis} shows and analyzes the results obtained by our tool, including the client and geographical distribution findings that this work presents.
In Section~\ref{sec:discussion} we discuss the possible hazards that the Eth2 network could face.
Finally, Section~\ref{sec:conclusion} concludes this work and presents some possible future directions.

\section{Background}
\label{sec:background}

Distributed applications among p2p networks have gained popularity with the pass of time over the last two decades. Starting in the 2000s, distributed file systems emerged with the launch of the first p2p network, Gnutella \cite{ripeanu2001peer}. Since back then, the idea of propagating information through the network by using p2p connections among peers has served as the base to develop modern distributed blockchain applications such as Bitcoin, Ethereum or Monero.

 Previous works \cite{matei2002mapping} \cite{sen2002analyzing} \cite{jovanovic2001modeling} pointed that decentralized p2p networks by nature tents to self-organize through the usage of methods like the Distributed Hash Tables (DHT) \cite{castro2002exploiting}, acquiring unique properties and layouts that can directly affect the performance, reliability, scalability, and sometimes anonymity of the applications that rely on them. They emphasized the importance that the topology of the overlay network had, proposing methods that would help to dig and discover the topology of the first p2p networks. 

As discussed in \cite{stutzbach2008characterizing}, the size of the p2p networks constantly grows as the size of the information that gets shared, leaving the previous approaches and methods outdated. In this approach, the authors included in the analysis non-intrusive techniques able to track: the available nodes over time, the traffic volume, and the host distribution in the network.

With the popularization of blockchain technology over the last decade, new distributed protocols based on blockchain technology appeared, e.g. Bitcoin and Ethereum. The impact of these decentralized protocols has augmented in the industrial and academic fields as shown in \cite{mohanta2019blockchain}, triggering remarkable attention on the p2p networks that they rely on. 

Previous works \cite{delgado2018cryptocurrency} \cite{deshpande2018btcmap} \cite{wang2020ethna} provide compelling insights into the actual p2p networks of the most prominent Blockchain protocols, Bitcoin and Ethereum. In these works, the authors show methods to track the number of nodes on the overlay, dig into the peer discovery over the network, and even some light message exchange of information among the peers. In this paper, we present the p2p network analysis over the Eth2 main net performed with our developed network crawler tool that we name Armiarma.

The proposed approach leverages some previous ideas in \cite{ben2018vivisecting}, \cite{kim2018measuring} and \cite{biryukov2014deanonymisation}, such as network crawlers as a tool to exploit possible network weaknesses that might be a hazard for the integrity of the protocol. As explained in those papers, the information about the connections between the nodes that form the network is as important as the information about the nodes them-self. In this approach, we generate a study on the new Eth2 p2p network, including the analysis of:
\begin{itemize}
    \item the discovered topology of the Eth2 p2p network overlay.
    \item the distribution between the client types that are available to participate in the network. 
    \item network traffic insights from the GossipSub protocol. 
\end{itemize}

\section{Eth2 network ecosystem}\label{sec:stateofart}

On a large decentralized blockchain that involves so many nodes and clients, the security and integrity of the protocol relies on the healthiness of the network. Being able to debug from inside all the occurring events, offers the possibility to get an in-depth view of the network status and might even help to prevent  performance issues as well as more serious network reliability problems or even security vulnerabilities. Escaping from the complex and tedious work of generating specific test cases for each of the clients, the community have tried to develop tools to interact easily with the network, without having to operate as a full client.

\subsection{Rumor}
\label{subsec:rumor}

Rumor is an interactive shell script tool for debugging and testing the interaction between the Eth2 clients. Originally started as a Eth2 networking tool written in Go, Rumor was designed as an alternative to become the common platform for testing the different Eth2 nodes in a less effort and time-consuming way. By offering a set of commands that can replicate most of the Eth2 network protocol interactions, Rumor offers the abstraction from the code implementation that makes setting a custom node, possible by simply tipping a few commands.

The platform is based on an environment close to the Bash shell that eases the test building labour, where some of the Bash syntax (e.g., variables and control flow statements) are accepted. Along with the scripting capabilities, Rumor includes a complex system of sub-environments (known in Rumor as actors) that allows to generate different nodes with different properties simultaneously. The different nodes distributed along the different actors share the common environment, which perform as link between them. This environment settles the possibility for the peers to exchange information, needed to perform a synchronization of actions based on constantly occurring events.
The governance over the node specs and its conduct makes Rumor the appropriate tool for testing and debugging the Eth2 clients and their behaviour under different conditions. The variety of commands that Rumor offers is still on a developing stage, but it already includes most of Eth2 networking specs such:

\begin{itemize}
    \item Generate a ENR identity for the node needed to participate on some of the Eth2 protocols.
    \item The Eth peer discovery protocol \emph{dv5.1}.
    \item Generate a peer database or \emph{peerstore} of the connected and discovered peers.
    \item Establish a connection with a given peer.
    \item Exchange Status and Metadata of the Beacon State with the connected peers.
    \item Listen and publish messages on the GossipSub p2p protocol used in Eth2.
    \item Send and respond to RPCs from and to the peers.
\end{itemize}

The different hosts configurations and behaviour strategies can be deployed in different ways. If the desired test is not too long, it can be launched from the \emph{rumor shell} by typing the commands in the desired order. For more repetitive tasks or tests, the platform offers the possibility to read the list of commands from a given file (e.g., \emph{script.rumor}). For more complex tasks, Rumor also offers the option to add a test-plan into a service, allowing the test-case to behave as a server.

\subsection{Node Discovery Protocol v5}
\label{subsec:dv5}

As a decentralized platform, Eth2 tries to avoid any point of centralization inside the proposed protocol. The networking area, and more specifically the peer discovery, is not an exception to this rule. The implemented GossipSub protocol, despite being a protocol oriented to a message propagation on decentralized applications, does not offer any kind of peer discovery service. Leaving that application in charge of this task. Eth2 developed its own node discovery protocol, Discovery 5 (\emph{dv5}), currently on its 5.1 version~\cite{dv5-protocol}. 

Dv5 focuses on the Ethereum Node Records (ENR) exchange along the peers. These node records are recorded in a DHT. This DTH offers the possibility to sample and search along the generated database, allowing to easily update the node record of a specific peer if modifications are detected. As entry point to the dv5 protocol, the nodes can start searching for others that publicly advertise themselves on specific topics, or as an alternative that improves the performance, a first connection with boot-nodes is possible.  

The peer discovery service is essential for the network in general. New peers joining the network need connections from where synchronize the chain. Peers that are already on the network also need to update from time to time their information source, ensuring that they are properly following the head of the chain. Furthermore, to increase the resilience of the network to Sybil attacks, dv5 provides a list of nodes on the network that the node could connect at any point.

\section{Methodology}
\label{sec:methodology}

The approach proposed in this paper tackles the lack of information about the performance of the GossipSub protocol and the peers on the Eth2 main net. The developed tool, Armiarma, is built on top of Rumor and it offers a simple yet powerful method that provides meaningful data about the p2p network and the Eth2 clients. The tool is divided into two parts, the Armiarma crawler and the Armiarma analyzer, as shown in figure~\ref{fig:ArmiarmaScheme} and described bellow.

\subsection{Armiarma Crawler}
\label{subsec:armiarmacrawler}

\begin{figure}[h]
    \centering
    \includegraphics[width=0.9\linewidth]{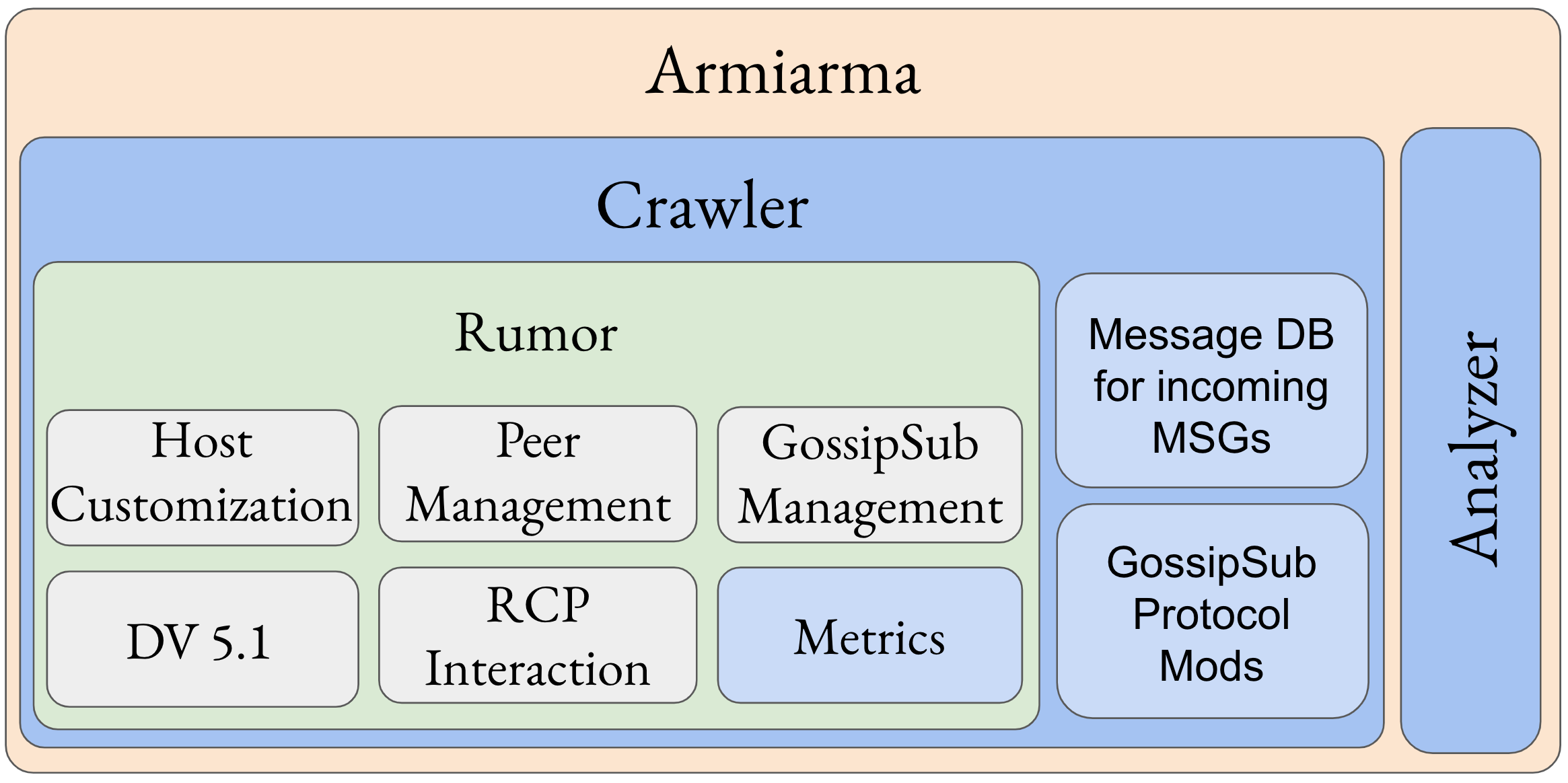}
    \caption{Scheme of Armiarma. Armimiarma additions to Rumor in blue.}
    \label{fig:ArmiarmaScheme}
\end{figure}

Armiarma Crawler is the data gatherer part of the tool. The crawler has been based on the Eth2 client debugging tool Rumor, and its performance is linked to its ability to discover and peer with nodes from the Eth2 network. The crawler is a specific use case of Rumor, described in Figure~\ref{fig:ArmiarmaScheme}. On top of Rumor, we build the custom host, that together with the modifications compiled in its repository \cite{armiarma-crawler}, provides a simplification of the data gathering, data processing and test reproducibility processes. The crawler can be directly launched from the tool, building autonomously the necessary environment to operate before it executes the chain of commands that will initialize  the custom host.
\hfill

The initialization process starts with the host set up assigning the IP and ports that will be used by the host. Followed by the ENR generation, the Eth2 ENR identity gets generated from the host info and by specifying the network that we want to belong to. Once we advertise on the ENR of the crawler that we participate on the Eth2 main net~\cite{main-net}, a Beacon Node is faked. Claiming to be at the genesis state prevents the crawler from being penalized for not offering services from the beacon chain, such as block synchronizations. To successfully fake an Eth2 client and being accepted by the rest of the peers, the custom host has been set up to be at the genesis state of the Beacon Chain.
After setting the host, the GossipSub p2p protocol is initialized, joining the main topics from the Eth2 specs. The PubSub protocol~\cite{libp2p-frok} Rumor GossipSub implementation has been modified~\cite{armiarma-crawler} to store all the data gathered from the peers we get connected to, getting from each of them:
\begin{enumerate}
    \item Information about the peer: Peer Id, Node Id, Client Type and Client Version, Pubkey, MultiAddress, IP, Country, City and Latency.
    \item The connection/disconnection events with the timestamp of each event.
    \item A Counter of every message we have received from each peer on the five main GossipSub topics of Eth2:
    \begin{itemize}
        \item BeaconBlock
        \item BeaconAggregateAndProof
        \item VoluntaryExit
        \item ProposerSlashing
        \item AttesterSlashing
    \end{itemize}
\end{enumerate}

Once the host is fully initialized, the peer discovering protocol \emph{dv5}~\cite{dv5-protocol} and the \emph{connectall} Rumor services are launched. This way, all the peers that we are able to find through the \emph{dv5} protocol are recorded into a peerstore and we attemp to connect with them. 
From all of those peers that we establish a connection, the crawler would start to get the messages from the GossipSub topics previously joined, generating as well the metrics previously mentioned. To save the metrics, a new fucntionality has been added, to exporting the recorded data into the previously defined \emph{test-project}.

\subsection{Armiarma Analyzer}
\label{subsec:armiarmaanalyzer}

While Armiarma crawler is focused on the interaction with the p2p network, recording into a database all the events and peer information, Armiarma analyzer gets in charge of performing an analysis on the raw information gathered. 
At the moment, the raw data parsing gets done by the crawler itself, exporting periodically the final metrics into a \emph{csv} file. This exported metrics compiles the individually processed data for each of the nodes. During this parsing process, the format of several items gets sorted or modified, aiming to ease further the data analysis. One of those format modifications corresponds to the count of the recorded connections and disconnections events, that serves to obtain the total connection time of each peer.
The analysis of the metrics gets performed in a subsequent moment when the user executes the commands on the tool. For that the Analyzer will work over the metrics from the given project-folder. In the analysis, the analyzer performs some filtering and analysis methods producing as result human-readable graphs. The obtained compilation of graphs and their insights will be discussed in section \ref{sec:analysis}.

\section{Analysis}
\label{sec:analysis}

In order to test the capabilities of the tool, we performed an experiment in which we let the crawler run for several days collecting as much data as possible.
The crawler has been set on a cloud server with limited resources, trying to replicate a low-computational power environment. The machine hosting the tool has 2 Intel Cores at 2.3GHz and 7GB of RAM, 50 GB of storage, and 250 Mbps of network bandwidth. The exact running period of the experiment is from 2021-06-19 to 2021-06-26. The discussion of the results obtained from the analysis has been divided into the three subsections described below.

\subsection{Eth2 Network Analysis}
\label{subsec:network-analysis}

In blockchain protocols, the network's security relies on a high number of nodes working together. The network has a margin of peers that can go offline without necessarily leading to the protocol crash. Eth2 needs at least 66\% of the validators actively attesting blocks to finalize~\cite{buterin2017casper}. 

\begin{figure*}[!htb]
    \minipage{0.32\textwidth}
        \includegraphics[width=1\linewidth]{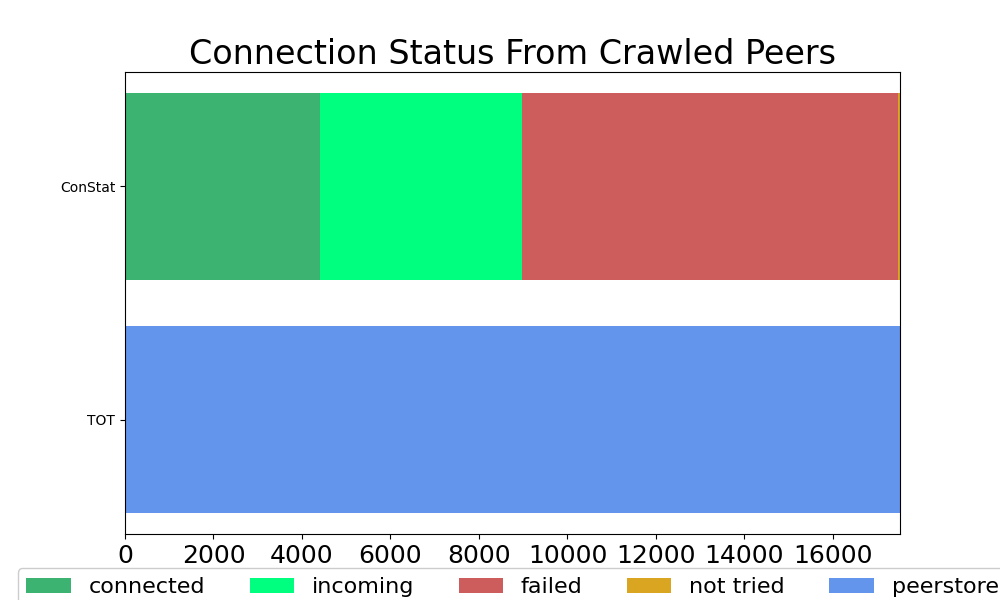}
        \caption{Connection status distribution of the discovered peers.}
        \label{fig:ConnStatusDistribution}
    \endminipage\hfill
    \minipage{0.32\textwidth}%
        \includegraphics[width=1\linewidth]{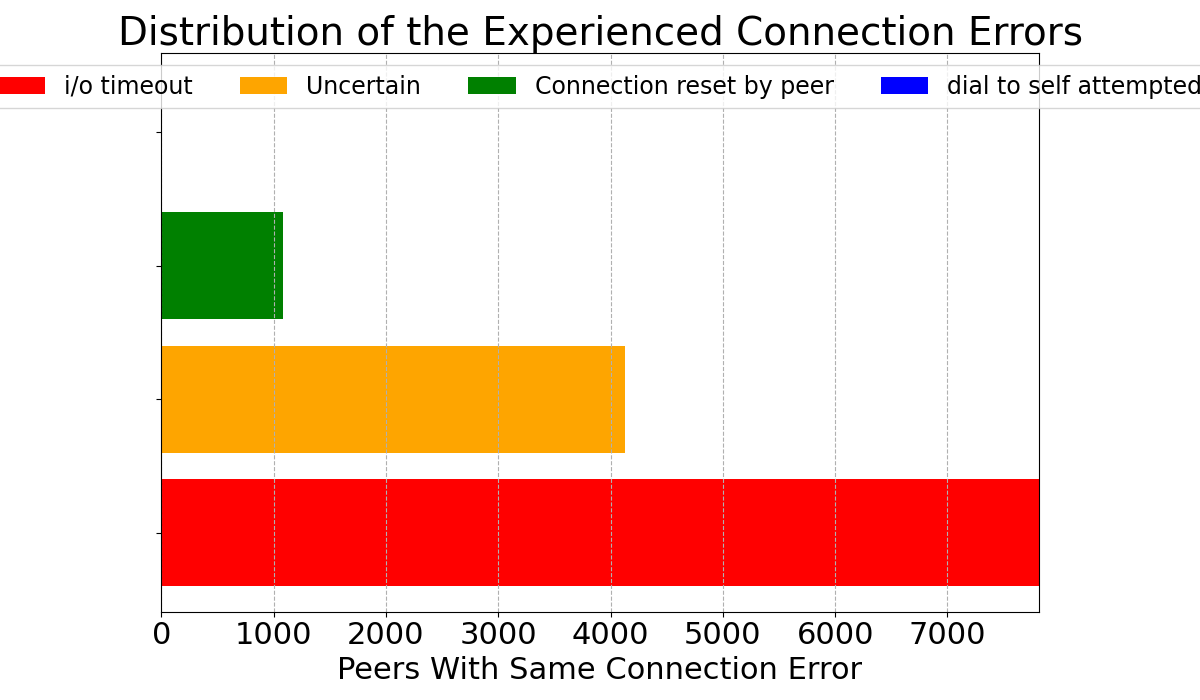}
        \caption{Error distribution on active connection attempts.}
        \label{fig:ErrorDistributions}
    \endminipage\hfill
    \minipage{0.32\textwidth}%
        \includegraphics[width=1\linewidth]{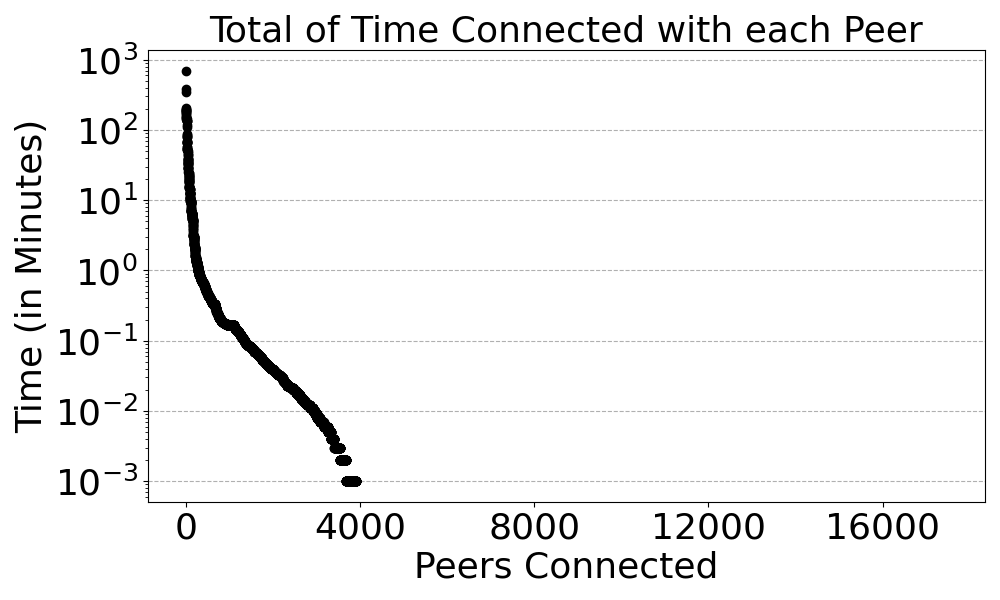}
        \caption{Amount of time connected with each peer.}
        \label{fig:TimeConnected}
    \endminipage
\end{figure*}

The first part of the analysis consists of having the crawler get a list of peers and connect to them. From the aggregation of $17516$ peers discovered by the \emph{dv5} protocol over the seven days of crawl, the crawler could connect with only $8980$ of them. 
We were expecting a higher connection rate. However, temporary disconnections or validators migrating from one client to another could not be the only reason for the non-connected peer ratio. The current implementation of the Eth2 main clients does not tent to keep the previous public keys for the beacon node whenever this one is rebooted. This choice increases the privacy of the validator hosted behind the node but also increases the chances of counting no longer active peers. It is possible then that the previously mentioned $8980$ connected peers measurement includes double-counted or rebooted peers.  The observed distribution could also be affected by a non-port forwarding configuration by some of the non-connected peers. Those peers stay under the protection of the ISP firewalls, preventing us from initiating the peering handshake. 

At the moment, we estimate that around $5000$ unique active nodes are participating in the network. 
However, from the peers we managed to connect with, only $4415$ of the successful connections were actively started by the crawler, while the other $4565$ of the peerings were incoming ones, as we can appreciate in Figure~\ref{fig:ConnStatusDistribution}. As previous works \cite{sen2002analyzing} \cite{castro2002exploiting} points out, the base of some p2p network it's likely to rely on a small portion of peers from the entire network. Often called \emph{core-peers} or \emph{super-peers}, these peers compound the nucleus of the network. These are peers that it is easy to connect with and transmit information reliably. The $4415$ peers we managed to connect and exchange data easily could represent a good portion of that \emph{nucleus} of the network. Previous work \cite{kiayias2017ouroboros} shows that in a network such as the Eth2 (based on PoS), peers must show stability. This is expected since the disconnection of validators leads to economic slashing. From now on, we will refer to those peers that show stability as \emph{steady-peers}.

Taking a closer look, in Figures \ref{fig:ErrorDistributions}, we can observe that from all the peers we tried to connect, the ones that failed reported different errors as a result of the dialing attempt. In the figure, we appreciate that 1081 of the peers reset the connection, $4132$ of them reported an uncertain error, while $7812$ of the peers reported a timeout error in the attempt to dial a connection. 

We can categorize two kinds of peers in this group: the active, reachable, and stable ones, and the active but not reachable peers. The first ones provided most of the messages that we saw. Meanwhile, the second type could reach the crawler via opportunistic requests and drop the connections once a reply is received. Furthermore, we note that the crawler does not fulfill all the functionalities of a full node. Therefore, requests for chain information or blocks are directly dropped from our side. Despite this apparent difference between the peer types, we also see in Figure~\ref{fig:TimeConnected} differences in the time we have been connected to the peers. These variations are related to the pruning process of the GossipSub protocol. This process aims to keep a predefined number of peer connections. Several client implementations reduce the maximum number of peers the node can connect to keep the computing power and network bandwidth low. 

\begin{figure*}[!htb]
    \minipage{0.32\textwidth}
        \includegraphics[width=1\linewidth]{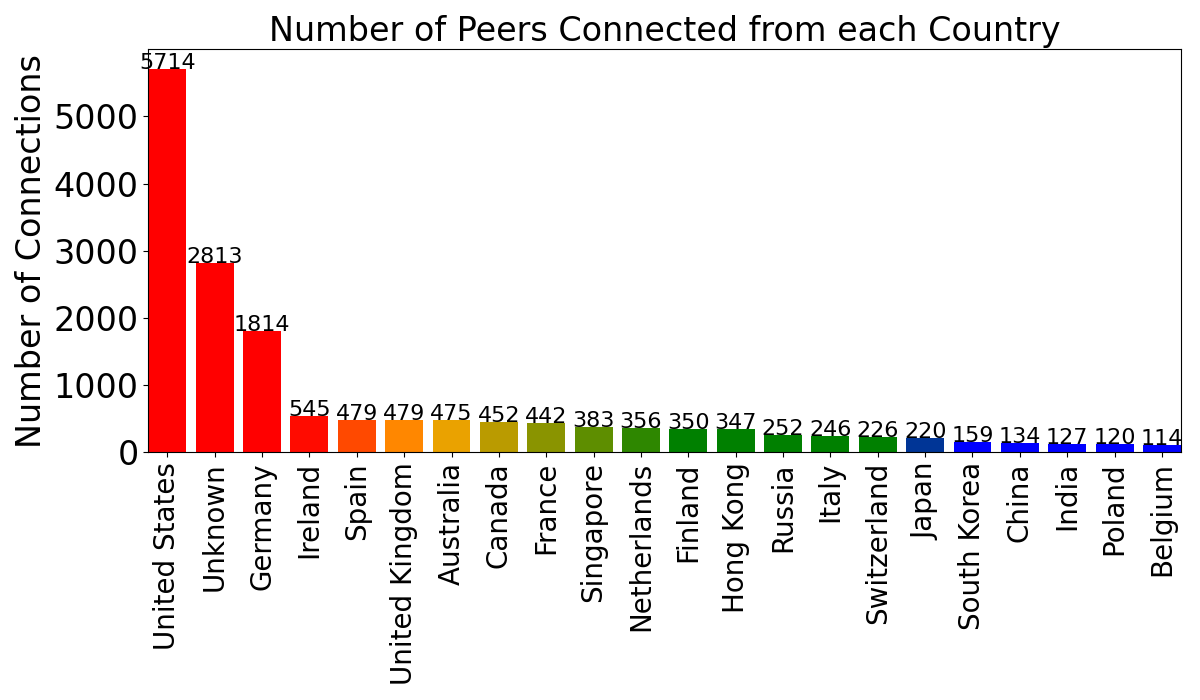}
        \caption{Number of peers classified by country of origin.}
        \label{fig:PeersPerCountries}
    \endminipage\hfill
    \minipage{0.32\textwidth}%
        \includegraphics[width=1\linewidth]{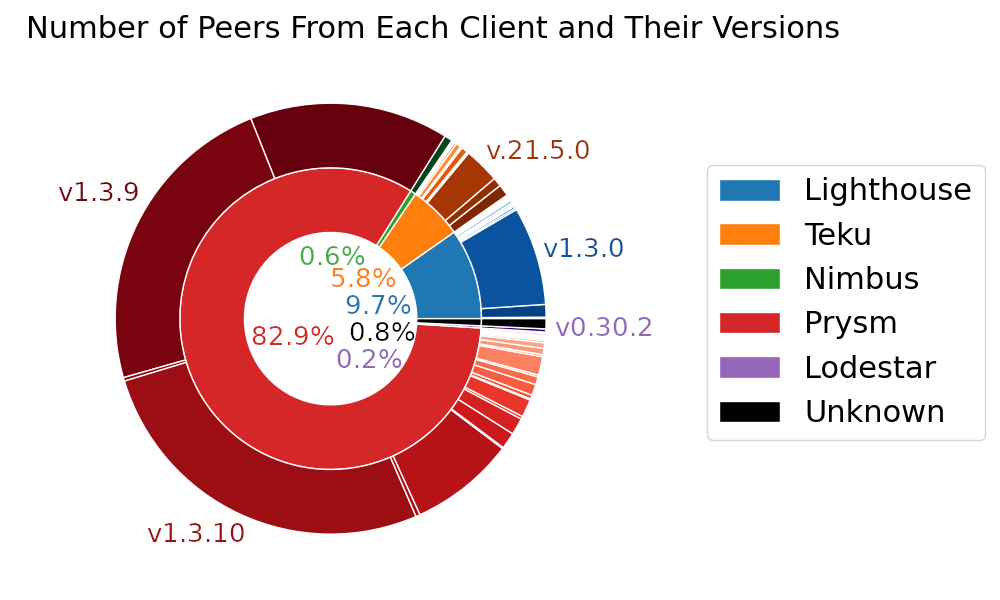}
        \caption{Number of peers connected from each client.}
        \label{fig:PeersPerClient}
    \endminipage\hfill
    \minipage{0.32\textwidth}%
        \includegraphics[width=1\linewidth]{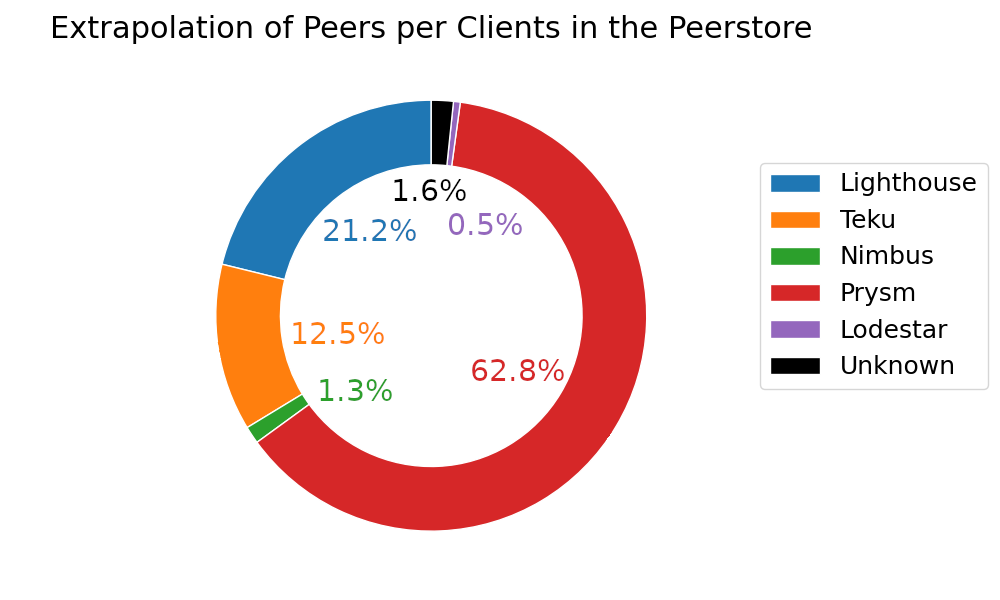}
        \caption{Extrapolation of the Client type distribution.}
        \label{fig:client-estimations}
    \endminipage
\end{figure*}

Blockchain networks are geographically distributed all around the world. Having the peers distributed means that the peers are less likely to get affected by the same type of regional problems or events (e.g., National censorship). In Figure~\ref{fig:PeersPerCountries}, we can appreciate that almost half of the peer distribution concentrates in two countries, the US (32,65\%) and Germany (10,37\%). Note that for visualization purposes the figure only displays countries with more than $100$ peers over the $97$ countries compiling the total list.

In a decentralized network such as Eth2, low latency connections are essential for tracking every occurring event as fast as possible. In this case, on the Eth2 Beacon Chain protocol, there is a new Slot event every $12$ seconds, which means that every $12$ seconds, a randomly chosen validator will have the chance to propose a Beacon Block and therefore receive the token reward. In Figure~\ref{fig:RTTWithPeers}, we can observe the Round Trip Time (RTT) of the latest connections that the crawler recorded while being connected to the peers. The RTT measures the time duration that it takes since we send a request to the destination peer until we receive a response. The latency between the destination peer and us is half of the RTT or lower. We can observe that the majority of peers stay with an RTT below three seconds\footnote{Note that some peers do not have any measured RTT assigned because the crawler didnot receive any answer from them.}, having an outlier with a $67.049$ seconds RTT. Note that the crawling process was done from a node located in Frankfurt and that the RTT includes the time the requester took to process the request. Thus, peers situated on the other side of the world, and peers with a higher work load would usually report a higher latency, e.g., the outlier peer with an IP from the Düsseldorf (See Section \ref{subsec:peer-analysis}).

\subsection{Eth2 Clients Interaction}
\label{subsec:client-analysis}
Currently, there are five Eth2 clients available to participate on the network. With our tool, we were able to detect which client is used by each connected peer, and it is even possible to get the version they are using, as shown in Figure~\ref{fig:PeersPerClient}. Note that \emph{unknown} peers could participate on the network as boot nodes, other crawlers, possible attackers, or just peers that did not report their client in the metadata. Also, we could see how many versions of each client are out there in the wild, as shown in Table~\ref{tab:client-versions}. This is useful because we can observe over a long period how client teams deploy new versions and how fast users update their clients.

As we can see in Figure~\ref{fig:PeersPerClient}, Prysm is the most connected client with 82.9\% of the connected peers ($7446$ peers). In contrast, in the Peerstore, we observe that 62.84\% of the peers use the TCP port used by Prysm clients ($13000$), showing that most nodes in the Eth2 network are using Prysm Clients, but in a lower percentage that the one \emph{observed} by the crawler.
\footnote{Although it is possible to change the dialing port of the host in all the clients, the percentage of users that change it to this one is low.} 
There are two main reasons for this disparity. First, the figure does not show all the peers the crawler \emph{discovered}, but only those it managed to exchange peer metadata with. 

Finally, issues related to port-forwarding also played a role in preventing some connections. To better understand how the client distribution represents the network, we calculated an estimation based on the advertised port and the observed distribution from those peers with whom we exchanged metadata. As shown in Figure~\ref{fig:client-estimations}, we can see that Prysm is the most dominant client in the network, followed by Lighthouse and Teku. This large disparity and lack of software decentralization is somehow concerning (See Section \ref{sec:discussion}). 

\begin{table}[ht]
  \caption{Number of versions observed of each client. }
  \label{tab:client-versions}
    \begin{center}
    \begin{tabular}{|c|c|c|c|c|c|} 
      \hline
      Lighthouse & Teku & Nimbus & Prysm & Lodestar & Unknown \\ 
      \hline	 
      17 & 17 & 1 & 29 & 1 & 1 \\
      \hline
    \end{tabular}
    \end{center}
\end{table}

\begin{figure*}[!htb]
    \minipage{0.32\textwidth}
        \includegraphics[width=1\linewidth]{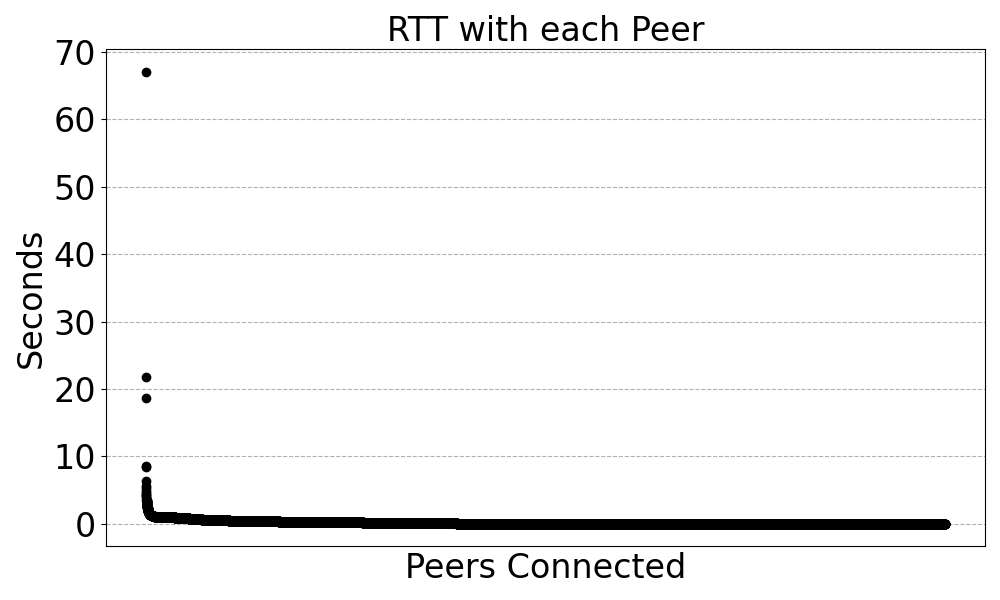}
        \caption{RTT distribution with peers.}
        \label{fig:RTTWithPeers}
    \endminipage\hfill
    \minipage{0.32\textwidth}%
        \includegraphics[width=1\linewidth]{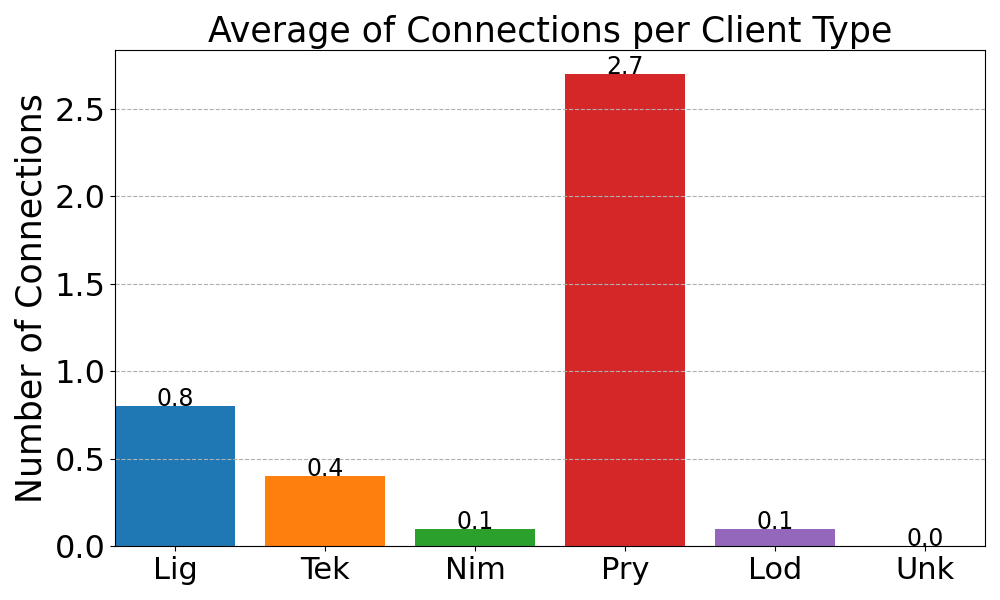}
        \caption{Average connections to peers classified by client types.}
        \label{fig:AverageOfConnectionsPerClientType}
    \endminipage\hfill
    \minipage{0.32\textwidth}%
        \includegraphics[width=1\linewidth]{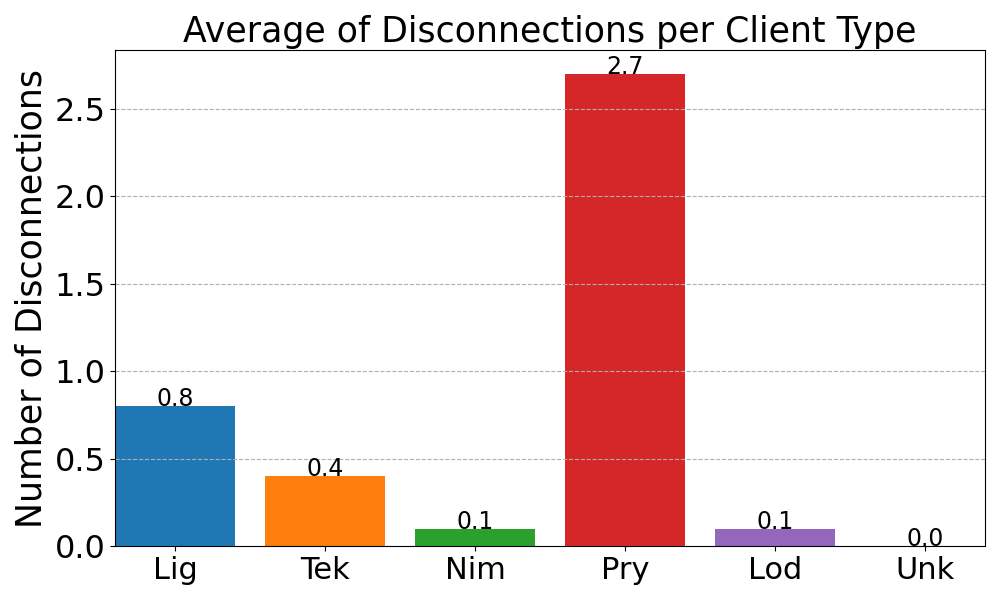}
        \caption{Average disconnections to peers classified by client.}
        \label{fig:AverageOfDisconnectionsPerClientType}   
    \endminipage
\end{figure*}

On the behavior comparison of the different clients with the crawler, we can observe several differences. First, we compare the average number of connections and disconnections for each one of the clients, as shown in Figures \ref{fig:AverageOfConnectionsPerClientType} and \ref{fig:AverageOfDisconnectionsPerClientType}. Please note that the tool was not launched, prioritizing the quality of the connections. It was prioritizing a larger number of connections from the Peerstore nodes. We can appreciate how the client distributions get mirrored in the number of dialed connections and disconnections in the figures. Prysm is the client type with a higher average in the perceived dialing events from those identified peers of the same client, followed by Lighthouse, Teku, Nimbus, and Lodestar in that order.


\begin{figure*}[!htb]
    \minipage{0.32\textwidth}
        \includegraphics[width=1\linewidth]{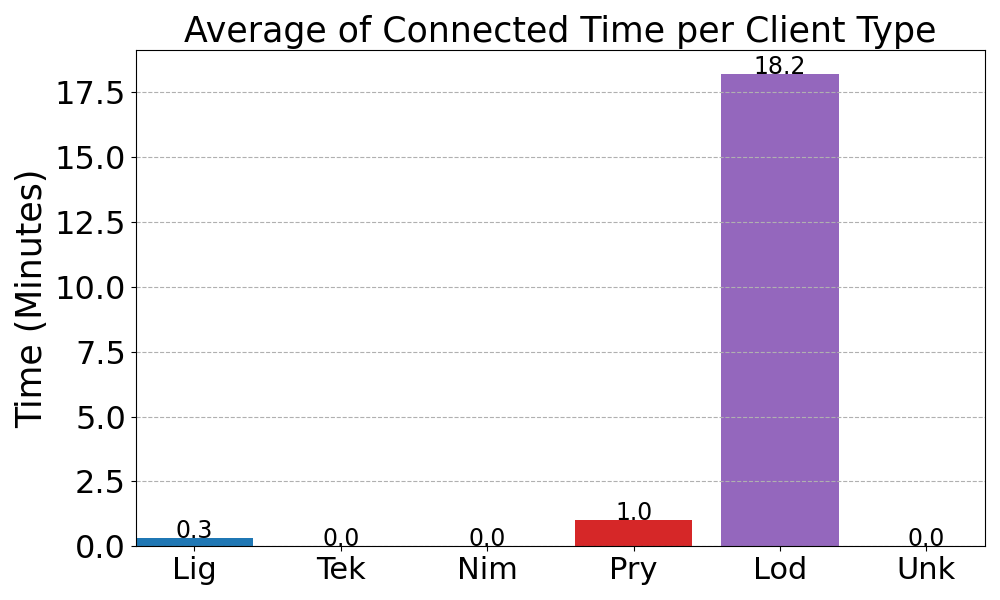}
        \caption{Average time connected to peers from each client.}
        \label{fig:AverageOfConnectedTimePerClientType}
    \endminipage\hfill
    \minipage{0.32\textwidth}
        \includegraphics[width=1\linewidth]{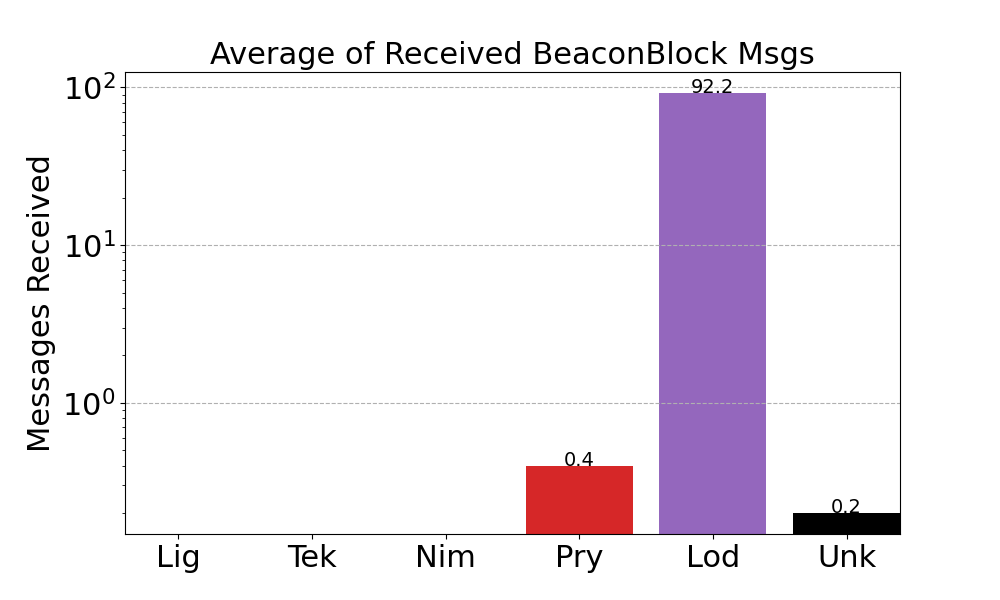}
        \caption{ Average of messages received on the BeaconBlock topic from peers filtered by clients.}
        \label{fig:MessageAverageFromBeaconBlock}
    \endminipage\hfill
    \minipage{0.32\textwidth}%
        \includegraphics[width=1\linewidth]{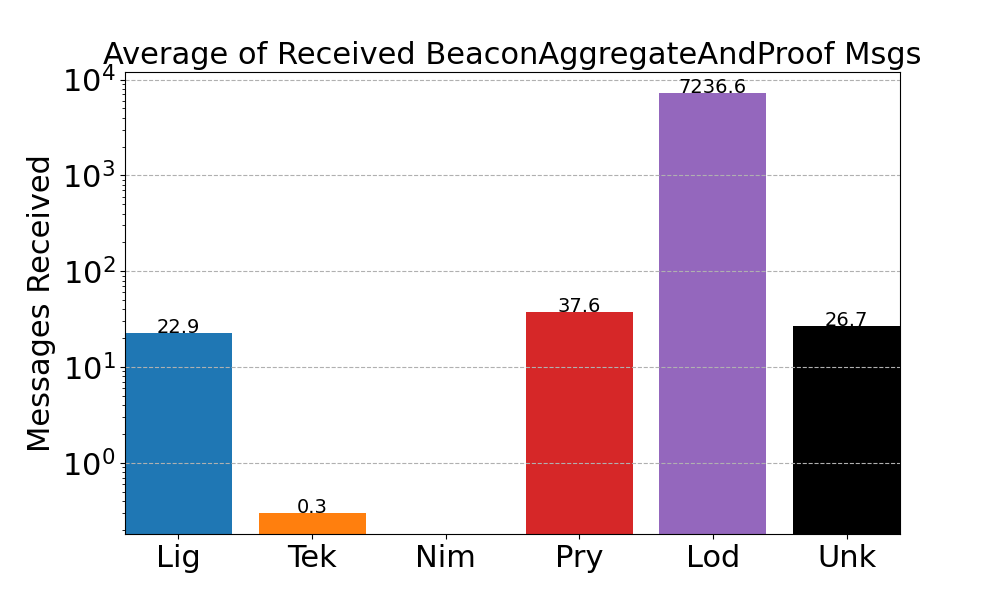}
        \caption{ Average of messages received on the BeaconAggregateAndProof topic from peers filtered by clients.}
        \label{fig:MessageAverageFromBeaconAggregateProof}
\endminipage
\end{figure*}

However, the number of connections does not equate to connected time because clients could attempt multiple connections while still disconnecting shortly after the connection has been established. Therefore, we keep track of both connections and disconnections, allowing us to have accurate total and average connection times. Thus, we plot in Figure~\ref{fig:AverageOfConnectedTimePerClientType} the average amount of time (in minutes) that each type of client was connected to our network monitoring tool.

We can observe that the results that Figures \ref{fig:AverageOfConnectionsPerClientType} and \ref{fig:AverageOfDisconnectionsPerClientType} do not match directly with the one observed in Figure~\ref{fig:AverageOfConnectedTimePerClientType}. While Prysm was showing a high level of activity on the dialing events, the average time spent connected to our crawler gets reduced to the second position. A similar relation can be observed in the other clients with the clear exception of Lodestar. The surprising difference between the perceived dialing events and the average connected time of Lodestar stands up. Despite having a low connection ratio, Lodestar peers showed a high level of stability on the performed connections. Several aspects could influence the observed phenomenon. 

First, we should point out the different peering strategies adopted by the clients while implementing the GossipSub protocol. While some clients can adopt more strict approaches to make sure their peers are contributing to pass information, others like Lodestar seem to have more flexible approaches, allowing for long connection periods.

In this case, we can see that the peering strategy we selected for the tool also added a bias to the averages. By prioritizing the number of connections over their quality, the tool drops the chance to connect to a stable peer if more unidentified peers are in the Peerstore. Thus, prioritizing for new peers instead of stable ones, plus the number of incoming connections, influences the connected time observed to each client.

On the other hand, computing the average of the metrics could also affect the observed results. The connections, disconnections, and the connected time get calculated from the aggregated connections and disconnections of all the nodes from a client type, divided by the number of nodes from the same type. The method used to obtain the averages leads to disperse any outstanding peer from the most connected client types. Meanwhile, for the lowest dialed client types, this average can get biased as it happens with Lodestar.

The combination of the low number of connection events and the low connected time on average for the Prysm nodes leads us to think that many Prysm nodes are behind non-forwarded ports or ISP firewalls. At the same time, it also leads us to believe that incoming connections tend to be more opportunistic than stable connections in the general term.

The time connected to a peer does provide some insight into the stability of the client. Some clients might be more aggressive than others pruning peers to avoid bad actors. However, this does not say much about the different clients' participation in the network. For that, we measure the total number of Beacon Block, and Aggregate and Proof messages received from each peer, grouping them by type of client.

\begin{figure*}[!htb]
\minipage{0.32\textwidth}
        \includegraphics[width=1\linewidth]{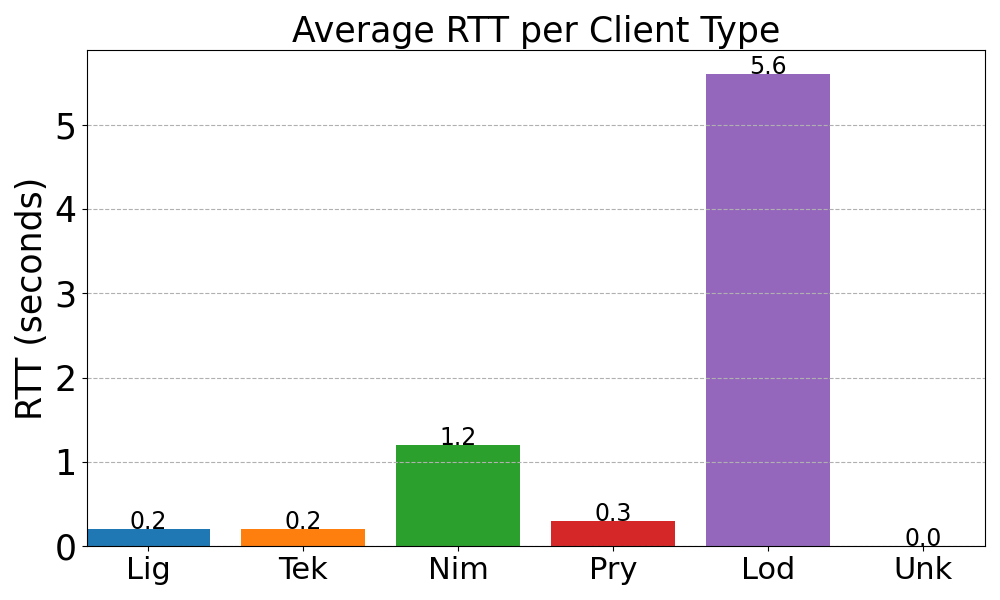}
        \caption{Average Round Trip Time (RTT) classified by client types.}\label{fig:AverageRTTPerClientType}
    \endminipage\hfill
    \minipage{0.32\textwidth}%
        \includegraphics[width=1\linewidth]{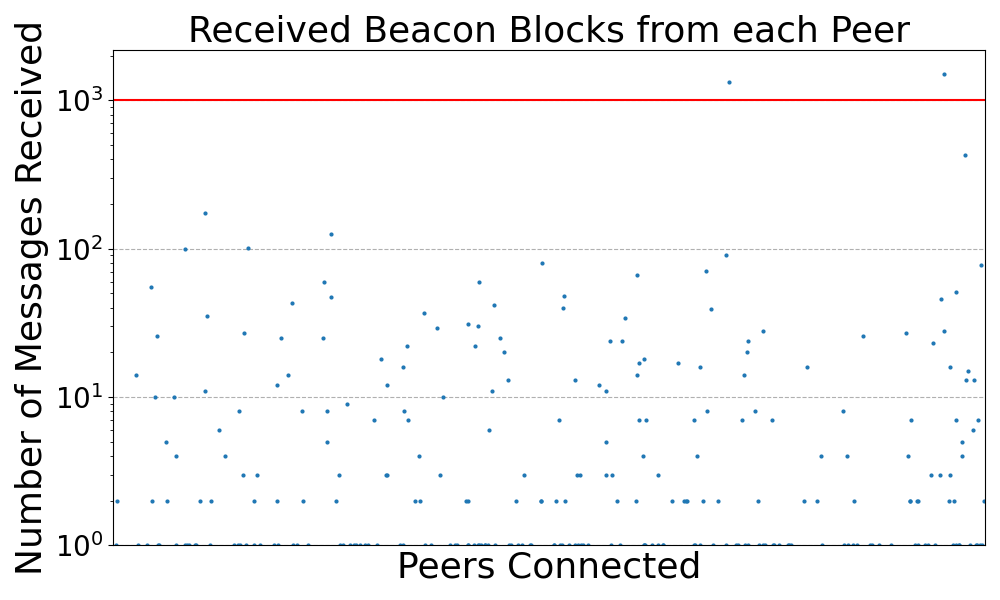}
        \caption{Number of Beacon Blocks Received from each Peer.}
        \label{fig:BeaconBlockMessagePerClient}
    \endminipage\hfill
    \minipage{0.32\textwidth}%
        \includegraphics[width=1\linewidth]{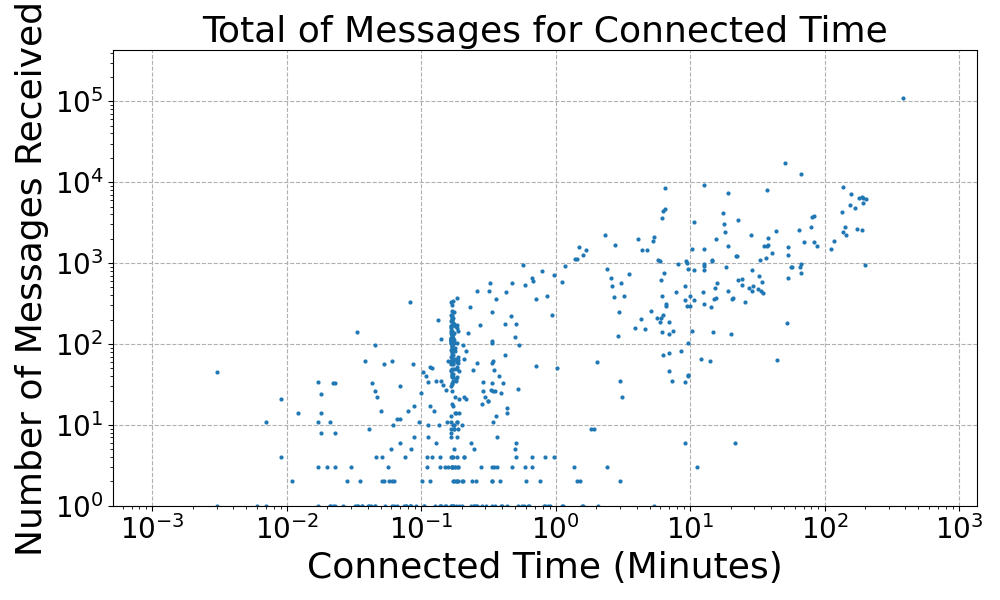}
        \caption{Total messages for Connected Time.}
        \label{fig:TotalMesagesPerTimeConnected}
\endminipage
\end{figure*}

As observed in Figures \ref{fig:MessageAverageFromBeaconBlock} and \ref{fig:MessageAverageFromBeaconAggregateProof} and as it was expected, the clients that in average have more stable connections tends to share more messages. In an attempt to understand a little better why Lodestar, Prysm, and Lighthouse are the most \emph{chatty} of the clients, we studied the average RTT for each client. In Figure~\ref{fig:AverageRTTPerClientType} we can appreciate that Nimbus got over x4 RTT and over x18 RTT for Lodestar with respect to the rest of the clients. As previously mentioned, peers like the Lodestar \emph{outlier} exposed in Table \ref{tab:peer-info} can significantly impact the average RTT of the clients, in particular when a client has only a few peers connected to the crawler. 


Leaving the type of client aside, we also wanted to see if all peers communicate in relatively similar amounts or if any node is communicating significantly more than others. Thus, we plot in Figure~\ref{fig:BeaconBlockMessagePerClient} the number of beacon Block messages received from each peer. 

One of the modifications that make this analysis unique is the modification that Armiarma implements on the GossipSub p2p protocol. The original protocol saves network bandwidth consumption by only sending the metadata of a message with nodes outside the mesh before sending the whole message. If the receiver has not seen the message before, then it asks for the entire message to be sent, if it has seen it before, then it does not ask for it. In addition to that, even for peers inside the mesh, if the message has already been seen, GossipSub does not forward this to the client application. Thus, future messages that have been seen from different peers will not be received. Our modification on the GossipSub protocol allows forwarding the message to the application even if the message was seen before, allowing the crawler to keep track of all the network traffic without increasing it.

In the results we can appreciate which nodes take the role of \emph{steady-peers}, situating themselves in the center of the overlay. As we can see, two peers communicated over 2000 Beacon blocks, while the large majority of the peers did not share more than 100. More precisely, 90\% of the BeaconBlocks were received only by 0.31\% of the peers, while 99.69\% of the peers transmitted less than 15 Beacon Blocks. The high number of network traffic perceived from just the 0.31\% of the network highlights how the Eth2 p2p network still shares the same network topology as the ones previously mentioned on Section~\ref{sec:background}. This topology also explains why some nodes are more chatty and share faster than others. 



In addition, we analyzed how many messages peers send concerning the time they keep connected. A node transmitting too many messages in a short connection time could signal a deny of service (DoS) attack. In contrast, a node with a very long connection but very few or no messages could signal an attempt of an eclipse attack. The distribution is shown in Figure~\ref{fig:TotalMesagesPerTimeConnected}. Note that in this figure, we count Beacon Blocks as well as Proof Aggregations messages. The distribution shows a dense cloud of peers that stayed connected between 0.01 and 1 minutes and transmitted between 0 and 1.000 messages. Particularly interesting is the case of the dot in the top right of the figure. This peer transmitted almost 230,000 messages and stayed connected for around 1.5 hours. We could call this a \emph{steady-peer} because it shows an extremely stable connection and participation on the beacon data transmission. 

\begin{table}[ht]
 \caption{Example of Gathered Information from a Peer}
 \label{tab:peer-info}
\begin{center}
\begin{tabular}{|r|l|} 
 \hline
 Data Type & Information \\ 
 \hline	
    Client Type & Lodestar \\
    Client Version & 0.30.2 \\
    Country & Germany \\
    City & Düsseldorf \\
    RTT & 67.049 \\
    Connections & 1 \\
    Disconnections & 1 \\
    Connected Time(min) & 86 \\
    Beacon Blocks & 430 \\
    Beacon Aggregations & 42508 \\
    Total Messages & 42938 \\
 \hline
\end{tabular}
\end{center}
\end{table}

\subsection{Individual Peer Analysis}
\label{subsec:peer-analysis}

The previous analysis performed on the Eth2 network shows a global view of the network. The gathered data can be used to analyze peers individually. We could extract a large amount of information regarding specific peers. To showcase an example, we show on Table~\ref{tab:peer-info} the data extracted from one of the peers in the network. We have omitted sensitive information for security and privacy reasons as peer Id, node Id, the public key, and IP related to the peer. This peer showed an unusually high RTT, which caught our attention. Combining the tree analysis that we presented can help us monitor and keep track of nodes suspected of malicious behavior. This demonstrates that the used techniques and the tool could be used for security purposes and the network status study.

\section{Discussion}
\label{sec:discussion}


This paper presents the observed geographical and client distribution in the recently launched Eth2 Beacon Chain. We found that although the platform incentivizes decentralization for security purposes, decentralization is not entirely achieved yet. The network crawler has identified that almost 43\% of the $17,516$ crawled nodes in the network are located in the US and Germany, representing the opposite of a geographically decentralized network. Previous works~\cite{kiffer2021under} have demonstrated that the networking performance completely relies on the location of the node. This means that nodes in regions far from the mentioned two countries could be in disadvantage, which could in turn lead to some level of geographical centralization.

In addition to the geographical distribution issue, the presented work displays that the network could be facing a critical mono-client dependency. The performed estimations show that about 62\% of the network relies on the same single client type, which could mean a finalization crisis if a vulnerability or bug is found in this particular client. This is one of the most critical points this crawler analysis highlights, because a bug impacted that specific client in April 24th causing finalization issues during multiple hours~\cite{prysm-bug}. 

We also analyzed the number of validators in relation to the number of Beacon nodes, and in relation to the number of IP addresses observed in the network. With $179,825$ active validators and $17,516$ crawled peers in Eth2 at the moment, we can estimate that there are around $10$ validators on average hosted behind each of the Beacon nodes. We also observe that from the $6635$ successfully identified peers (37\% of the peerstore),  $530$ share the same IP address with at least one more node. In fact, we observed that 100 of the identified IPs host more than 4 Beacon nodes at the same time. So far, the analysis represents that 1.5\% of the identified IPs host 9.8\% of the identified validators. This again, points to a level of decentralization lower than the ideal one targeted with the transition from PoW to PoS. 

Overall, this crawler analysis highlights several concerning points for the recently launched Eth2 Beacon Chain. From several perspectives, geographical-wise, implementation-wise as well as the physical hosting of the Eth2 validators, we have observed a not ideal level of decentralization. The target goal was that anyone, anywhere with a low-power device, such as Raspberry Pi, could participate in the Eth2 network. This is technically feasible today, however in practice we observe a completely different picture. One reason for this could be that the network is still very young and Eth2 has not been fully deployed yet. However, we believe it is important to boost the efforts to guarantee a much higher level of decentralization. 


\section{Conclusion And Future Work}
\label{sec:conclusion}

In this paper, we have presented some of the results we extracted from the Eth2 p2p network analysis. The collected information, transparently gathered with our network monitoring tool Armiarma, allowed us to get a solid overview of the network state. Our results show the healthy behavior in the network but also some concerns regarding decentralization. 

The introduced study identified certain geographical, client, and possible IP/Entities centralization degrees that could mean a potential hazard at any point of the networks' performance. This work intends to report the inconsistencies that the network is facing on its path towards this new Eth2.

Even though the presented analysis provided meaningful network insights about the Ethereum2 networks' topology, we are aware of the current limitations that the Armiarma tool has. Our future plan is to extend the analysis with a deeper track of the received messages, optimize the currently used methods to achieve a larger number of identified peers, and upgrade the tool towards a distributed crawling strategy that would allow us to perceive more accurate peer information.


\section*{Acknowledgment}

This work has been supported by the Ethereum Foundation under Grant FY20-0198. We would like to thank the researchers of the Ethereum Foundation for their feedback and suggestions. In particular, we would like to thank @Protolambda as the main Rumor maintainer for his help setting up our experiments and his constructive feedback on this study.  

\bibliographystyle{IEEEtran}
\bibliography{armiarma}

\end{document}